\documentclass[aps,pra,reprint]{revtex4-1}
\usepackage{graphicx}
\usepackage{dcolumn}
\usepackage{bm}
\usepackage{hyperref}
\usepackage{amssymb}
\usepackage{amsmath}
\usepackage{epsf}
\usepackage[caption=false]{subfig}
\usepackage{epstopdf}
\usepackage{amsfonts}
\usepackage{color}

\usepackage{pst-all}
\usepackage{graphicx}
\begin{document}
\title{Statistics of work done in degenerate parametric amplification process}
\author{Hari Kumar Yadalam and Upendra Harbola}
\affiliation{Department of Inorganic and Physical Chemistry, Indian Institute of
Science, Bangalore, 560012, India.}
\begin{abstract}
We study statistics of work done by two classical electric field pumps (two-photon and one-photon resonant pumps) 
on a quantum optical oscillator.  
We compute moment generating function for the energy change of the oscillator,
interpreted as work done by the classical drives on the quantum oscillator
starting out in a thermalized Boltzmann state. 
The moment generating function is inverted, analytically when only one of the pumps is turned on and 
numerically when both the pumps are turned on, to get the probability
function for the work. The resulting probability function for the work done by
the classical drive  is shown to satisfy transient detailed and integral work fluctuation theorems. 
Interestingly, we find that, in order for the 
work distribution function to satisfy the fluctuation theorem  in presence of both the drivings, relative phase of 
drivings need to be shifted by $\pi$, this is related to the broken time reversal 
symmetry of the Hamiltonian. 
\end{abstract}
\maketitle
\section*{Introduction}
Work done by external forces on isolated mesoscopic systems, unlike their
macroscopic counterparts \cite{Callen2013}, is
subject to fluctuations \cite{Esposito2009, Campisi2011, Seifert2012} due to the
smallness of the system size. 
These fluctuations could be due to the uncertainty of the initial state and due
to the quantum nature of evolution and measurement process. 
Despite the noisy nature of the work done by the external force on the nanoscale
system, these fluctuations exhibit 
marvelous symmetry property which links the frequency of certain amount of work
done by the external force on the 
system to the frequency that the same amount of work is extracted by the
external drive. Further, these symmetry 
properties of probability function for work done by the external force on the
system, termed as work-fluctuation-theorems, 
are one of the first class of fluctuation theorems discovered for out of
equilibrium systems 
\cite{Jarzynski1997, Jarzynski2004}. 
This fluctuation relation states that, for a driven system, the probability that an external force extracts
certain amount of work from the system 
is finite but exponentially suppressed compared to the probability that 
exactly the same amount of 
work is performed on the system. 
In this sense fluctuation theorems promote the inequality of
second law of thermodynamics (for the 
dissipated work) to an equality \cite{Boksenbojm2010}. 
These results are universal in the sense that only ingredients that are
sufficient to establish these relationships are
the equilibrium canonical nature of the initial state and the microscopic
reversibility of the underlying dynamics, they
are insensitive to the nature of microscopic details of the system
\cite{Esposito2009,Campisi2011,Seifert2012}. 
Nevertheless, the probability function for work done by external force on the
system is not universal and depends on 
the microscopic details of the system.  Work distribution function has been
computed for a variety of situations for 
both classical \cite{Seifert2012,Klages2013} and quantum systems
\cite{Talkner2008,Deffner2008,Deffner2010,Yi2011,Ford2012,Quan2012,Ngo2012,
Sotiriadis2013,Fei2014,Leonard2015,Brunelli2015,Jarzynski2015,Lobejko2017,
Blattmann2017,Mata2017,Wang2018}. Experimental measurement of 
work distribution and subsequent demonstration of Jarzynski-Crooks fluctuation
theorems for classical systems is well established
\cite{Liphardt2002,Collin2005,Douarche2005,Blickle2006,Ciliberto2010,Seifert2012,Saira2012,Klages2013}. 
While for quantum systems, there is no work operator, it was initially
confusing to define work in quantum case
\cite{Allahverdyan2005,Talkner2007,Campisi2011q,Campisi2011}. Subsequently, a
two-point measurement protocol was proposed 
\cite{Kurchan2000,Tasaki2000,Talkner2007,Campisi2011q,Campisi2011} to define
work in a single realization. This was crucial for  proving quantum versions of
fluctuation theorems \cite{Esposito2009,Talkner2007,Campisi2011q,Campisi2011}.
It has been challenging to implement two-point measurement protocol
experimentally. 
However there have been some recent theoretical proposals of experiments
\cite{Huber2008,Heyl2012,Dorner2013,Mazzola2013,Talarico2016} and
implementations \cite{Batalhao2014,An2015,De2018,Smith2018} of work measurements
in quantum systems either directly through two-point measurement 
protocol or indirectly. 

 In this work we consider, two-point measurement protocol to study work
statistics in generating displaced squeezed thermal states of quantum optical oscillator 
by driving the optical oscillator starting in Gibbs state by two classical pumps resonant with two-photon and one-photon transition. 
The scenario where only two photon resonant pump is on corresponds to the standard degenerate parametric amplification process. 
 This work is motivated by recent interesting  proposals
\cite{Huang2012,Rossnagel2014,Abah2014,Alicki2015,Niedenzu2016,Manzano2016,Klaers2017,Agarwalla2017,Niedenzu2017,Niedenzu2018,Manzano2018} of using 
 squeezed thermal reservoirs in quantum heat engines to surpass standard Carnot
efficiency. Squeezed thermal states of light
\cite{Lvovsky2015,Schnabel2017,Barsotti2018} can 
 be realized using well established parametric
amplification process
\cite{Louisell1961,Gordon1963,Mollow1967a,Mollow1967b,Raiford1974,Mivsta1978,Wodkiewicz1983,Mark1984,Crouch1988,Mertens1993,Galve2009,Acosta2015,Andersen2016}. 
We assume that the oscillator is isolated from the 
environment during the driving process. We thus interpret the change in the 
energy of the oscillator as work performed by the classical drives.  
We compute the work-distribution function for this process. 
The work distribution function is shown to satisfy quantum version of Jarzynski-Crooks  fluctuation theorem
\cite{Jarzynski1997,Crooks1998,Crooks1999,Monnai2005,Kurchan2000,Tasaki2000,Talkner2007,Campisi2011q}. 
We note that work statistics has been studied in generating squeezed thermal state of a harmonic oscillator 
in Refs. \cite{Deffner2008,Deffner2010}, where classical driving is modeled through 
temporal modulation of harmonic oscillator frequency and analytic results
for work statistics were obtained approximately under limiting conditions. In Ref. \cite{Talkner2008}, work statistics has been studied in 
generating a displaced thermal  state of a harmonic oscillator. In this work we consider a general process where a quantum optical 
oscillator is driven by two classical pumps, one resonant with the two-photon transition and other resonant with the one-photon transition. 
We obtain exact result for the moment generating function for work and show that interference between the two drivings affects the work statistics 
in a non-trivial way and plays important role in satisfying the fluctuation theorem. 
In the limit when only one of the pumps is on, we analytically invert the moment generating functions to get work distribution function. 
Our results can be tested experimentally using classical optical
simulation of quantum dynamics as proposed 
in Ref. \cite{Talarico2016} and experimentally implemented in Ref.
\cite{De2018}.
 
 In the next section, we describe the model system and obtain
generating function for work done within the two-point
 measurement scheme and verify fluctuation theorems for work. The generating function is inverted to compute the
probability distribution function in 
 Sec. (\ref{prob-dist}) and cumulants of work are analyzed. We conclude in Sec. (\ref{conc}).
 
\section{Generating function for the work}
\label{sec-1}

We consider a quantum optical process of driving an isolated optical oscillator by two classical pumps \cite{Scully1997}.
We assume that one of the classical pumps has the same frequency as that of the optical oscillator ($\frac{\epsilon}{\hbar}$) 
and the second pump has twice the frequency of the optical oscillator. 
We treat the coupling between classical pumps and quantum optical oscillator within rotating wave approximation. 
We assume that the initial state of the optical oscillator
is thermal state (i.e., oscillator is kept in 
contact with thermal reservoir till time zero).
Note that the driving is non-adiabatic with respect to the 
natural frequency of the oscillator.
The Hamiltonian describing the evolution of quantum optical oscillator driven by
classical pumps (written in the interaction picture)
\cite{Scully1997,Agarwal2013,Garrison2014} is,
\begin{eqnarray}
\label{eq-1}
 \hat{H} &=& i\hbar \left[z_{1}^{} b_{}^{\dag} - z_{1}^{*}b_{}^{}\right]
 +  \frac{i\hbar}{2} \left[z_{2}^{} b_{}^{\dag} b_{}^{\dag} - z_{2}^{*}b_{}^{}b_{}^{}\right],
\end{eqnarray}
where $z_{1}^{}/z_{2}^{}$ is the product of coupling constant and electric field amplitude of one-photon/two-photon resonant classical pump and
$b_{}^{}/b_{}^{\dag}$ are the annihilation/creation operator for the quantum optical oscillator. 
We interpret the energy change of the quantum oscillator as the work done
by the classical pumps 
on it \cite{Jarzynski2007,Horowitz2007,Talkner2007,Campisi2011q}. With these
assumptions, the work done by the classical drives is proportional to the number of quanta ($n \in \mathbb{Z}$) of energy exchanged with the quantum oscillator. 
The probability distribution for the number of quanta exchanged by the classical pumps with the quantum optical oscillator during time period '$t$' 
is given as,
\begin{eqnarray}
\label{eq-2}
 P[n ; t] &=& \int_{0}^{2\pi}\frac{d\chi}{2\pi} \mathcal{Z}[\chi
; t] e_{}^{- i \chi n},
\end{eqnarray}
with $\mathcal{Z}[\chi ; t]$ being the moment generating function for the work
done, obtained within the two-point 
measurement protocol
\cite{Esposito2009,Campisi2011,Kurchan2000,Tasaki2000,Monnai2005}:  
the optical oscillator is kept in contact with a thermal
reservoir till time $0$,
it is then disconnected from the reservoir and the first energy measurement is made.
It is then subjected to driving by two classical pumps till
time $t$. At time $t$, classical drives are
turned off and the second energy measurement is made. The differences in the
energies measured initially and finally is the work done in a single
realization. With this two-point measurement protocol, 
$\mathcal{Z}[\chi ; t]$ is given as, 
\begin{eqnarray}
\label{eq-3}
 \mathcal{Z}[\chi ; t] &=&
\mathbf{Tr}\left[\mathcal{U}_{\chi}^{}(t,0)\rho_{}^{}(0)\mathcal{U}_{-\chi}^{}(0
,t)\right],
\end{eqnarray}
with $\rho_{}^{}(0) = \frac{e_{}^{-\beta \epsilon b_{}^{\dag}
b_{}^{}}}{\mathbf{Tr}
\left[e_{}^{-\beta \epsilon b_{}^{\dag} b_{}^{}}\right]}$ is the initial density
matrix of the optical oscillator 
which is assumed to be at the thermal state, and
$\mathcal{U}_{\chi}^{}(t_{1}^{},t_{2}^{})=e_{}^{-\frac{i}{\hbar}}
\hat{H}_{\chi}^{}(t_{1}^{}-t_{2}^{})$ is the twisted evolution operator in the
interaction picture with 
\begin{eqnarray}
\label{eq-4}
\hat{H}_{\chi}^{}= i\hbar \left[z_{1}^{} e_{}^{i \frac{ \chi}{2}}
b_{}^{\dag} - z_{1}^{*} 
e_{}^{-i \frac{ \chi}{2}} b_{}^{}\right] + \frac{i\hbar}{2} \left[z_{2}^{} e_{}^{i  \chi}
b_{}^{\dag} b_{}^{\dag} - z_{2}^{*} 
e_{}^{-i  \chi} b_{}^{}b_{}^{}\right].\nonumber\\
\end{eqnarray}

In order to compute $\mathcal{Z}[\chi ; t]$, it is convenient to work with the
Weyl generating function, $\mathcal{G}[\zeta_{}^{},\zeta_{}^{*} ; t]$,  defined (in the interaction picture) as 
\cite{Scully1997,Carmichael2009,Carmichael2003},
\begin{eqnarray}
\label{eq-5}
 \mathcal{G}_{\chi}^{}[\zeta_{}^{},\zeta_{}^{*} ; t] &=& \mathbf{Tr}
 \left[e_{}^{ i \left[ \zeta_{}^{*} b_{}^{} + \zeta_{}^{} b_{}^{\dag} \right]} 
 \mathcal{U}_{\chi}^{}(t,0)\rho_{}^{}(0)\mathcal{U}_{-\chi}^{}(0,t) \right],
\end{eqnarray}
and then, 
$ \mathcal{Z}[\chi ; t] = \mathcal{G}_{\chi}[0,0; t]$.
 Using standard techniques from quantum optics literature
\cite{Scully1997,Carmichael2003,Carmichael2009},  
 it can be shown that (using Eqs. (\ref{eq-3}) $\&$ (\ref{eq-4}))$\mathcal{G}_\chi [\zeta,\zeta^* ; t]$ satisfies the
following evolution equation,
 \begin{eqnarray}
 \label{eq-6}
  &&\frac{\partial}{\partial t} \mathcal{G}_{\chi}^{}[\zeta_{}^{},\zeta_{}^{*} ;
t] =\nonumber\\
&&\left[\frac{1}{2} \begin{pmatrix} \zeta_{}^{*} \\ \zeta_{}^{} \\
\frac{\partial}{\partial \zeta_{}^{*}} \\ \frac{\partial}{\partial \zeta_{}^{}}
\end{pmatrix}_{}^T \begin{pmatrix} \mathcal{A} & \mathcal{B} \\ \mathcal{B}^{T}_{} & \mathcal{C} \end{pmatrix} \begin{pmatrix} \zeta_{}^{*} \\
\zeta_{}^{} \\ \frac{\partial}{\partial \zeta_{}^{*}} \\
\frac{\partial}{\partial \zeta_{}^{}} \end{pmatrix}+\begin{pmatrix} d_{1}^{} \\ d_{2}^{} \end{pmatrix}_{}^{T}\begin{pmatrix} \zeta_{}^{*} \\
\zeta_{}^{} \\ \frac{\partial}{\partial \zeta_{}^{*}} \\
\frac{\partial}{\partial \zeta_{}^{}} \end{pmatrix}\right]\mathcal{G}_{\chi}^{}[\zeta_{}^{},\zeta_{}^{*} ; t],\nonumber\\
 \end{eqnarray}
where $V^T$ represents transpose of a vector $V$ and,
\begin{eqnarray}
\label{eq-7}
 \mathcal{A} &=& 
 \begin{pmatrix}  
 - i\frac{ z_{2}^{} }{2}\sin( \chi_{}^{}) & 0 \\ 
  0 & - i \frac{ z_{2}^{*} }{2}\sin( \chi_{}^{}) 
  \end{pmatrix}\nonumber\\
  \mathcal{B}_{}^{} &=&   \begin{pmatrix}  
  0 & z_{2}^{} \cos( \chi_{}^{}) \\  
  z_{2}^{*} \cos( \chi_{}^{}) & 0 
  \end{pmatrix}\nonumber\\
  \mathcal{C}_{}^{} &=& 
   \begin{pmatrix}  
 - 2 i z_{2}^{*}\sin( \chi_{}^{}) & 0 \\ 
  0 & - 2 i z_{2}^{} \sin( \chi_{}^{}) 
      \end{pmatrix},
\end{eqnarray}

\begin{eqnarray}
\label{eq-8}
d_{1}^{}&=&\begin{pmatrix}i z_{1}^{}\cos(\frac{\chi}{2}) & i z_{1}^{*}\cos(\frac{\chi}{2})\end{pmatrix}_{}^{T}\nonumber\\
 d_{2}^{}&=&\begin{pmatrix}2 z_{1}^{*}\sin(\frac{\chi}{2}) &  2 z_{1}^{}\sin(\frac{\chi}{2})\end{pmatrix}_{}^{T}.
\end{eqnarray}

The above parabolic partial differential equation, Eq. (\ref{eq-6}), has to
be solved along with the 
initial condition $\mathcal{G}[\zeta,\zeta^* ; t]\Big|_{t=0}^{} =
\mathbf{Tr}\left[e^{ i [ \zeta^* b 
+ \zeta b^\dag] }\rho(0)\right] = e^{-\frac{1}{2} \begin{pmatrix} \zeta^* \\
\zeta_{}^{}
\end{pmatrix}^{T} D \begin{pmatrix} \zeta_{}^{*} \\ \zeta_{}^{} \end{pmatrix}}$
with $2\times 2$ matrix $D$ is 
defined as, $D_{ij}= (1-\delta_{ij})(f+\frac{1}{2})$ where, $f=(e^{\beta
\epsilon}-1)^{-1}$.  
 The solution of Eq. (\ref{eq-6}) is obtained as (see appendix), 
\begin{eqnarray}
\label{eq-9}
 \mathcal{G}_{\chi}^{}[\zeta_{}^{},\zeta_{}^{*} ; t] &=& \int_{\bar\zeta \in 
 \mathbb{C}}^{} d_{}^{2} \bar\zeta \mathbf{G}_{\chi}^{}[\zeta_{}^{} ,
\zeta_{}^{*} 
 ; t | \bar\zeta_{}^{} , \bar\zeta_{}^{*} ; 0 ]
\mathcal{G}_{}^{}[\bar\zeta_{}^{},\bar\zeta_{}^{*} ; 0] ,
\end{eqnarray}
with
\begin{widetext}
\begin{eqnarray}
\label{eq-10}
 &&\mathbf{G}_{\chi}^{}[\zeta_{}^{} , \zeta_{}^{*} ; t | \bar\zeta_{}^{} ,
\bar\zeta_{}^{*} ; 0 ] 
 =\nonumber\\
 &&\frac{e_{}^{\int_{0}^{t}dt_{1}^{}\left[{\mathcal{U}_{21}^{}(t_{1}^{})}_{}^{T}d_{1}^{}-{\mathcal{U}_{11}^{}(t_{1}^{})}_{}^{T}d_{2}^{}\right]_{}^{T}\int_{t_{1}^{}}^{t}dt_{2}^{}\left[{\mathcal{U}_{22}^{}(t_{2}^{})}_{}^{T}d_{1}^{}-{\mathcal{U}_{12}^{}(t_{2}^{})}_{}^{T}d_{2}^{}\right]}}{\pi \sqrt{-\det[\mathcal{U}_{21}(t)]}}
  e_{}^{-\frac{1}{2} 
 \begin{pmatrix} 
 \zeta_{}^{*} \\ 
 \zeta_{}^{} 
 \end{pmatrix}_{}^{T} {\mathcal{U}_{12}^{}(t)}_{}^{}{\mathcal{U}_{22}^{}(t)}_{}^{-1} 
\begin{pmatrix} \zeta_{}^{*} \\ 
\zeta_{}^{} 
\end{pmatrix}}e_{}^{\begin{pmatrix} \bar\zeta_{}^{*}
\\ 
\bar\zeta_{}^{} \end{pmatrix}_{}^{T}\int_{0}^{t}dt_{1}^{}\left[{\mathcal{U}_{22}^{}(t_{1}^{})}_{}^{T}d_{1}^{}-{\mathcal{U}_{12}^{}(t_{1}^{})}_{}^{T}d_{2}^{}\right]}\nonumber\\
&&e_{}^{-\frac{1}{2} \left[\begin{pmatrix} \zeta_{}^{*} \\ 
\zeta_{}^{} \end{pmatrix}-{\mathcal{U}_{22}^{}(t)}_{}^{}\left(\begin{pmatrix} \bar\zeta_{}^{*}
\\ 
\bar\zeta_{}^{} \end{pmatrix}+\int_{0}^{t}dt_{1}^{}\left[{\mathcal{U}_{21}^{}(t_{1}^{})}_{}^{T}d_{1}^{}-{\mathcal{U}_{11}^{}(t_{1}^{})}_{}^{T}d_{2}^{}\right]\right)\right]_{}^{T}
\left({\mathcal{U}_{21}^{}(t)}_{}^{}{\mathcal{U}_{22}^{}(t)}_{}^{T}\right)_{}^{-1}
\left[\begin{pmatrix} \zeta_{}^{*} \\ 
\zeta_{}^{} \end{pmatrix}-{\mathcal{U}_{22}^{}(t)}_{}^{}\left(\begin{pmatrix} \bar\zeta_{}^{*}
\\ 
\bar\zeta_{}^{} \end{pmatrix}+\int_{0}^{t}dt_{1}^{}\left[{\mathcal{U}_{21}^{}(t_{1}^{})}_{}^{T}d_{1}^{}-{\mathcal{U}_{11}^{}(t_{1}^{})}_{}^{T}d_{2}^{}\right]\right)\right]_{}^{}},\nonumber\\
\end{eqnarray}
where $\begin{pmatrix} \mathcal{U}_{11}^{}(t) & \mathcal{U}_{12}^{}(t) \\ 
\mathcal{U}_{21}^{}(t) & \mathcal{U}_{22}^{}(t) \end{pmatrix}=
e_{}^{\begin{pmatrix} \mathcal{B} & -\mathcal{A} \\  \mathcal{C}^{}_{} & -\mathcal{B}_{}^{T} \end{pmatrix}t}$.
\end{widetext}
Using Eq. (\ref{eq-9}) along with Eq. (\ref{eq-10}) in  $ \mathcal{Z}[\chi ; t] = \mathcal{G}_{\chi}[0,0; t]$ and performing $\bar\zeta$ integrals, we get,
\begin{widetext}
\begin{eqnarray}
\label{eq-11}
 \mathcal{Z}[\chi ; t] &=& \frac{1}{\sqrt{\det\left[{\mathcal{U}_{22}^{}(t)}_{}^{}+{\mathcal{U}_{21}^{}(t)}_{}^{}D\right]}}\nonumber\\
 &&e_{}^{\int_{0}^{t}dt_{1}^{}\int_{t_{1}^{}}^{t}dt_{2}^{}\left[{\mathcal{U}_{21}^{}(t_{1}^{})}_{}^{T}d_{1}^{}-{\mathcal{U}_{11}^{}(t_{1}^{})}_{}^{T}d_{2}^{}\right]_{}^{T}D\left[{\mathcal{U}_{21}^{}(t)}_{}^{}D+{\mathcal{U}_{22}^{}(t)}_{}^{}\right]_{}^{-1}{\mathcal{U}_{21}^{}(t)}_{}^{}\left[{\mathcal{U}_{22}^{}(t_{2}^{})}_{}^{T}d_{1}^{}-{\mathcal{U}_{12}^{}(t_{2}^{})}_{}^{T}d_{2}^{}\right]_{}^{}}\nonumber\\
 &&e_{}^{-\int_{0}^{t}dt_{1}^{}\int_{0}^{t_{1}^{}}dt_{2}^{}\left[{\mathcal{U}_{21}^{}(t_{1}^{})}_{}^{T}d_{1}^{}-{\mathcal{U}_{11}^{}(t_{1}^{})}_{}^{T}d_{2}^{}\right]_{}^{T}\left[\mathbb{I}_{2 \times 2}^{}-D\left[{\mathcal{U}_{21}^{}(t)}_{}^{}D+{\mathcal{U}_{22}^{}(t)}_{}^{}\right]_{}^{-1}{\mathcal{U}_{21}^{}(t)}_{}^{}\right]\left[{\mathcal{U}_{22}^{}(t_{2}^{})}_{}^{T}d_{1}^{}-{\mathcal{U}_{12}^{}(t_{2}^{})}_{}^{T}d_{2}^{}\right]_{}^{}}\nonumber\\
 &&e_{}^{\frac{1}{2}\int_{0}^{t}dt_{1}^{}\int_{0}^{t}dt_{2}^{}\left[{\mathcal{U}_{22}^{}(t_{1}^{})}_{}^{T}d_{1}^{}-{\mathcal{U}_{12}^{}(t_{1}^{})}_{}^{T}d_{2}^{}\right]_{}^{T}\left[{\mathcal{U}_{21}^{}(t)}_{}^{}D+{\mathcal{U}_{22}^{}(t)}_{}^{}\right]_{}^{-1}{\mathcal{U}_{21}^{}(t)}_{}^{}\left[{\mathcal{U}_{22}^{}(t_{2}^{})}_{}^{T}d_{1}^{}-{\mathcal{U}_{12}^{}(t_{2}^{})}_{}^{T}d_{2}^{}\right]_{}^{}}\nonumber\\
 &&e_{}^{-\frac{1}{2}\int_{0}^{t}dt_{1}^{}\int_{0}^{t}dt_{2}^{}\left[{\mathcal{U}_{21}^{}(t_{1}^{})}_{}^{T}d_{1}^{}-{\mathcal{U}_{11}^{}(t_{1}^{})}_{}^{T}d_{2}^{}\right]_{}^{T}D\left[{\mathcal{U}_{21}^{}(t)}_{}^{}D+{\mathcal{U}_{22}^{}(t)}_{}^{}\right]_{}^{-1}{\mathcal{U}_{22}^{}(t)}_{}^{}\left[{\mathcal{U}_{21}^{}(t_{2}^{})}_{}^{T}d_{1}^{}-{\mathcal{U}_{11}^{}(t_{2}^{})}_{}^{T}d_{2}^{}\right]_{}^{}}.
\end{eqnarray}
\end{widetext}
The final expression for the moment generating function for work can obtained by using explicit expressions for ${\mathcal{U}_{xy}^{}(t)}_{}^{}$. This gives,
\begin{widetext}
\begin{eqnarray}
\label{eq-12}
 \mathcal{Z}[\chi_{}^{},\phi_{}^{}; t]&=&
 \frac{e_{}^{\frac{|z_{1}^{}|_{}^{2}\left(\frac{\text{sinh}(|\frac{z_{2}^{}}{2}|t)}{|\frac{z_{2}^{}}{2}|}\right)_{}^{2}\left[\left(1+f\right)_{}^{}
 \left(e_{}^{ i  \chi}-1\right)+f_{}^{} \left(e_{}^{-i   \chi}-1\right)\right]\left[\text{cosh}(|z_{2}^{}|t)+\text{cos}(\phi_{}^{})\text{sinh}(|z_{2}^{}|t)\left[\left(1+f\right)_{}^{} e_{}^{ i \chi}-f_{}^{} e_{}^{-i \chi}\right]\right]}{1-\text{sinh}^2(|z_{2}^{}| t)\left[\left(1+f\right)_{}^{2}
 \left(e_{}^{ i 2  \chi}-1\right)+f_{}^{2} \left(e_{}^{-i 2  \chi}-1\right)\right]}}}{\sqrt{1-\text{sinh}^2(|z_{2}^{}| t)\left[\left(1+f\right)_{}^{2}
 \left(e_{}^{ i 2  \chi}-1\right)+f_{}^{2} \left(e_{}^{-i 2  \chi}-1\right)\right]}},
\end{eqnarray}
\end{widetext}
where $\phi=2\text{Arg}(z_{1}^{})-\text{Arg}(z_{2}^{})$.

It is clear that 
$\mathcal{Z}[\chi_{}^{},\phi_{}^{}; t] \neq \mathcal{Z}[-\chi_{}^{}+i\beta_{}^{}\epsilon_{}^{},\phi_{}^{}; t]$, which is related to the broken time-reversal 
symmetry of the Hamiltonian \cite{Esposito2009,Campisi2011}, i.e, $\hat{H}_{}^{} \neq \mathcal{T}\hat{H}_{}^{}\mathcal{T}_{}^{-1}$ ($\mathcal{T}$ is the time-reversal operator). 
To recover Jarzynski-Crooks-Buchkov-Kuzolev fluctuation theorem, 
work distribution for time forward and backward trajectories need to be compared. 
The backward evolution \cite{Esposito2009,Campisi2011} is governed by $\mathcal{T}\hat{H}_{}^{}\mathcal{T}_{}^{-1}=-i\hbar \left[z_{1}^{*} b_{}^{\dag} - z_{1}^{}b_{}^{}\right]
 -  \frac{i\hbar}{2} \left[z_{2}^{*} b_{}^{\dag} b_{}^{\dag} - z_{2}^{}b_{}^{}b_{}^{}\right] \neq \hat{H}_{}^{}$. In order to recover the time-reversal symmetry, 
 we also need to change $\phi$ to $\pi-\phi$. This leads to the Gallavotti-Cohen symmetry for the work moment 
 generating function : $\mathcal{Z}[\chi_{}^{},\phi_{}^{}; t] = \mathcal{Z}[-\chi_{}^{}+i\beta_{}^{}\epsilon_{}^{},\pi-\phi_{}^{}; t]$. 
 This, in turn, leads to the transient 
 work fluctuation theorem : $\frac{ P[+n,\phi ; t]}{ P[-n, \pi-\phi ; t]} = e_{}^{ \beta  \epsilon n } \Rightarrow  \langle e_{}^{-\beta \epsilon n} \rangle = 1$.
But for cases where only one of the pumps is present (i.e., $z_{1}^{}=0$ or $z_{2}^{}=0$), $\mathcal{Z}[\chi_{}^{},\phi_{}^{}; t]$ becomes independent of $\phi_{}^{}$. 
This is because, for the 
case when only one of the pumps is present, the phase of the electric field can be gauged out (for the initial thermal state) and doesn't appear 
in the expression for $\mathcal{Z}[\chi_{}^{},\phi_{}^{}; t]$. When both the pumps are present, the phases of both the pumps cannot be gauged out. 
Hence work statistics is unaffected by the phase
of the classical fields for the cases $z_{1}^{}=0$ or $z_{2}^{}=0$ which is similar to the result noted in Ref.
\cite{Talkner2008} where the phases of the classical field is shown to have no
influence on the work statistics during the process of coherent displacement of harmonic oscillator from a thermal state. 
For the cases when only one of the pumps is present, the moment generating function and hence the probability distribution function for the work 
for time forward and backward processes is the same. 

In the next section, we discuss the cases $z_{1}^{}=0$ and $z_{2}^{}=0$ separately and present work distribution for the each case. 

 \section{Statistics of the work done}
\label{prob-dist}
It is convenient to represent $P[n ; t]$, defined in Eq. (\ref{eq-2}) along with Eq. (\ref{eq-12}) as a contour integral around unit circle in 
complex plane \cite{Arfken2013} as, 
\begin{widetext}
\begin{eqnarray}
\label{eq-13}
  &&P[n ; t] =\nonumber\\
  &&\oint_{|\xi|=1}\frac{d\xi}{2\pi i \xi^{n+1}} 
\sqrt{\frac{\left(1-\frac{1}{{\xi_{+}^{}(0)}_{}^{2}}\right)\left(1-{\xi_{-}^{}(0)}_{}^{2}\right)}{
\left(1-\frac{\xi_{}^{2}}{{\xi_{+}^{}(0)}_{}^{2}}\right)\left(1-\frac{{\xi_{-}^{}(0)}_{}^{2}}{\xi_{}^{2}}
\right)}}~e^{\alpha(\phi,t)\left(\xi_{}^{}-1\right)\frac{\left(1+\frac{{\xi_{-}^{}(\phi)}_{}^{}}{\xi_{}^{}}\right)
\left(1+\frac{\xi_{}^{}}{{\xi_{+}^{}(\phi)}_{}^{}}\right)\left(1+\frac{{\xi_{-}^{}(\phi)}_{}^{}{\xi_{+}^{}(\phi)}_{}^{}}{\xi_{}^{}}\right)}{
\left(1+{\xi_{-}^{}(\phi)}_{}^{}\right)\left(1+\frac{1}{{\xi_{+}^{}(\phi)}_{}^{}}\right)
\left(1+{\xi_{-}^{}(\phi)}_{}^{}{\xi_{+}^{}(\phi)}_{}^{}\right)}\frac{\left(1-\frac{1}{{\xi_{+}^{}(0)}_{}^{2}}\right)
\left(1-{\xi_{-}^{}(0)}_{}^{2}\right)}{\left(1-\frac{\xi_{}^{2}}{{\xi_{+}^{}(0)}_{}^{2}}\right)\left(1-\frac{{\xi_{-}^{}(0)}_{}^{2}}{\xi_{}^{2}}
\right)}},\nonumber\\
\end{eqnarray}
\end{widetext}
where 
\begin{eqnarray}
\label{eq-14}
&&\alpha(\phi,t)=\nonumber\\
&&|z_{1}^{}|_{}^{2}\left(\frac{\text{sinh}(|\frac{z_{2}^{}}{2}|t)}{|\frac{z_{2}^{}}{2}|}\right)_{}^{2}
\left[\text{cosh}(|z_{2}^{}|t)+\text{cos}(\phi)\text{sinh}(|z_{2}^{}|t)\right],\nonumber\\
\end{eqnarray}
and
\begin{eqnarray}
\label{eq-15}
{\xi_{\pm}^{}(\phi)}_{}^{}&=&\frac{1 \pm \sqrt{1 + 4 f \left(1+f\right)\text{cos}_{}^{2}(\phi)\text{tanh}^2(|z_{2}^{}|t)}}{2 \left(1+f\right)\text{cos}_{}^{}(\phi)\text{tanh}(|z_{2}^{}| t)}.
\end{eqnarray}

For the case $z_{1}^{}\neq 0 \neq z_{2}^{}$, we were not able to invert the moment generating function analytically to get probability function for work. 
Below we present analytical results for $z_{1}^{}=0$ and $z_{2}^{}=0$ cases and then discuss numerical results for the general case.

\subsection{$z_{2}^{}=0$ case}
Taking $z_{2}^{} \rightarrow 0$ limit of Eq.(\ref{eq-12}), the moment generating function of work is,
\begin{eqnarray}
\label{eq-16}
 \mathcal{Z}[\chi_{}^{}; t] &=& e_{}^{|z_{1}^{}|_{}^{2}t_{}^{2}\left[\left(1+f\right)\left(e_{}^{i \chi}-1\right)+f\left(e_{}^{-i \chi}-1\right)\right]}
\end{eqnarray}
This moment generating function corresponds to a bi-poissonian stochastic process. The above expression for $ \mathcal{Z}[\chi_{}^{}; t]$ 
is a special case (resonant drive) of more general expression for work generating function derived in Ref. \cite{Talkner2008} 
for the general displacement drive.
Cumulant generating function for work, $\ln\mathcal{Z}[\chi_{}^{}; t]$, clearly scales quadratically with time ($t_{}^{}$) and hence all cumulants (obtained using $(-i)_{}^{n}\left(\frac{d}{d \chi}\right)_{}^{n}\ln\mathcal{Z}[\chi_{}^{}; t]|_{\chi=0}^{}$) scale as $t_{}^{2}$. First two cumulants of work are :
$\langle n \rangle=|z_{1}^{}|_{}^{2}t_{}^{2}$ and $\langle\left( n - \langle n \rangle\right)_{}^{2}\rangle=|z_{1}^{}|_{}^{2}t_{}^{2}\left(1+2f\right)$.
Probability function of work can be obtained analytically by converting Fourier inversion to integral over unit circle in complex plane (Eq.(\ref{eq-13})) as :
\begin{eqnarray}
\label{eq-17}
 P[n ; t] &=& \oint_{|\xi|=1}^{}\frac{d\xi}{2\pi i}\frac{1}{\xi_{}^{n+1}} e_{}^{|z_{1}^{}|_{}^{2}t_{}^{2}\left[\left(1+f\right)\left(\xi_{}^{}-1\right)+f\left(\frac{1}{\xi_{}^{}}-1\right)\right]}
\end{eqnarray}
Observing that the function $e_{}^{|z_{1}^{}|_{}^{2}t_{}^{2}\left[\left(1+f\right)\left(\xi_{}^{}-1\right)+f\left(\frac{1}{\xi_{}^{}}-1\right)\right]}$ is an analytic function of complex variable $\xi_{}^{}$ 
in the entire complex plane except at $\xi_{}^{}=0,\infty$  where it has essential singularity (hence it is analytic in the center punctured unit disk), 
the above integral gives Laurent series (around $\xi_{}^{}=0$) coefficients of 
$e_{}^{|z_{1}^{}|_{}^{2}t_{}^{2}\left[\left(1+f\right)\left(\xi_{}^{}-1\right)+f\left(\frac{1}{\xi_{}^{}}-1\right)\right]}$. This gives ,
\begin{eqnarray}
\label{eq-18}
 P[n ; t] &=& \left(\frac{1+f}{f}\right)_{}^{\frac{n}{2}}\frac{|z_{1}^{}|_{}^{2|n|} t_{}^{2|n|}\left[f \left(1+f\right)\right]_{}^{|n|/2}}{\Gamma[1+|n|]}e_{}^{-|z_{1}^{}|_{}^{2} t_{}^{2}\left(1+2f\right)}\nonumber\\
 &&\mathbb{F}_{1}^{0}\left[1+|n|;f \left(1+f\right) |z_{1}^{}|_{}^{4} t_{}^{4}\right],
\end{eqnarray}
where the generalized hyper-geometric function of variable $x_{}^{}$ is defined as $\mathbb{F}_{n}^{m}\left[a_{1}^{},\cdots,a_{m}^{},b_{1}^{},\cdots,b_{n}^{};x\right]=\sum_{k=0}^{\infty}\frac{\left(a_{1}^{}\right)_{k}^{}\cdots\left(a_{m}^{}\right)_{k}^{}}{\left(b_{1}^{}\right)_{k}^{}\cdots\left(b_{n}^{}\right)_{k}^{}}\frac{x_{}^{k}}{\Gamma[k+1]}$ 
(here Pocchammer symbol is $\left(c_{}^{}\right)_{r}^{}=\frac{\Gamma[c_{}^{}+r]}{\Gamma[c_{}^{}]}$) \cite{Milne1972,Arfken2013,Gradshteyn2013}. 
It is important to note that probability function for work takes same form for both the forward and backward processes. Using this explicit expression for $P[n ; t]$, 
the detailed work fluctuation theorem and hence the integral fluctuation theorem can be verified. 

The probability that no work is performed on the system ($P[0;t]$) is given as,
\begin{eqnarray}
\label{eq-19}
 P[0;t]&=& e_{}^{-|z_{1}^{}|_{}^{2} t_{}^{2}\left(1+2f\right)} I_{0}^{}[\sqrt{f(1+f)}|z_{1}^{}|_{}^{2}t_{}^{2}]
\end{eqnarray}
where $I_{0}^{}[z]$ is modified Bessel function of first kind of order 0 of variable $z$. In the zero temperature limit, 
$\beta \epsilon \rightarrow \infty \Rightarrow f \rightarrow 0$,  $P[0;t] = e_{}^{-|z_{1}^{}|_{}^{2} t_{}^{2}}$, 
which means, if the oscillator's initial state is the ground state, number of microscopic realizations where no net 
work is performed by the displacement/linear drive on the oscillator decays with time as a Gaussian function. 
Further, in the zero temperature limit
\begin{eqnarray}
\label{eq-21}
 P[n;t]&=&\frac{|z_{1}^{}|_{}^{2n}t_{}^{2n}}{\Gamma[1+n]}e_{}^{-|z_{1}^{}|_{}^{2} t_{}^{2}} \Theta[n],
\end{eqnarray}
where $\Theta[n]=1\ \text{iff}\ n\geq0\ \text{else}\ \Theta[n]=0$. This means, if the oscillator's initial state is ground state, there are no microscopic realizations where work is extracted by the classical drive from the oscillator. 
Further the term $t_{}^{2n}$ competes with the exponential decay ($e_{}^{-|z_{1}^{}|_{}^{2} t_{}^{2}}$), shifting the value of most probable work ($n$) to higher values for larger times. 

\subsection{$z_{1}^{}=0$ case}
For $z_{1}^{}=0$ case, the moment generating function for work can be obtained by taking $z_{1}^{} \rightarrow 0$ limit of Eq.(\ref{eq-12}). This gives,
\begin{eqnarray}
\label{eq-22}
 &&\mathcal{Z}[\chi_{}^{}; t]=\nonumber\\
 &&\frac{1}{\sqrt{1-\text{sinh}^2(|z_{2}^{}| t)\left[\left(1+f\right)_{}^{2}
 \left(e_{}^{ i 2  \chi}-1\right)+f_{}^{2} \left(e_{}^{-i 2  \chi}-1\right)\right]}}.\nonumber\\
 \end{eqnarray}
This leads to an average work given by, $\langle n \rangle = \left(1+ 2
f\right)\text{sinh}^2(|z_{2}^{}| t)\geq 0$. 
Thus, on average, work is done on the 
quantum oscillator and grows exponentially in time. This is to be contrasted with the quadratic time-dependence for the $z_{2}^{}=0$ case 
discussed above. Second cumulant of work
distribution is 
$\langle\left( n - \langle n \rangle\right)_{}^{2}\rangle=\left(1+\left(1+2f\right)_{}^{2}
\text{cosh}(2|z_{2}^{}| t)\right)\text{sinh}^2(|gz| t) \geq 0$. 
This indicates that the distribution function becomes broader exponentially with
time. Similar to the previous section, the probability function for the work is expressed as the complex contour integral (Eq.(\ref{eq-13})) as,
\begin{eqnarray}
\label{eq-23}
P[n ; t] &=&
\oint_{|\xi|=1}^{}\frac{d\xi}{2\pi i}
\frac{1}{\xi_{}^{n+1}} 
\sqrt{\frac{\left(1-\frac{1}{{\xi_{+}^{}(0)}_{}^{2}}\right)\left(1-{\xi_{-}^{}(0)}_{}^{2}\right)}
{\left(1-\frac{\xi_{}^{2}}{{\xi_{+}^{}(0)}_{}^{2}}\right)\left(1-\frac{{\xi_{-}^{}(0)}_{}^{2}}{\xi_{}^{2}}\right)}}\nonumber\\
\end{eqnarray}
Noticing $|{\xi_{-}^{}(0)}_{}^{}|<1<|{\xi_{+}^{}(0)}_{}^{}|$ (for $0 < |z_{2}^{}| t < \infty$ and $0 < f
< \infty$), we observe that the function with square-root in the above integrand 
is a multivalued complex function with four branch points at $\pm{\xi_{-}^{}(0)}_{}^{}$ and $\pm {\xi_{+}^{}(0)}_{}^{}$. 
A single valued branch can be chosen
for this function by defining the branch cut as union of two straight lines
joining $-{\xi_{-}^{}(0)}_{}^{}$ with $+{\xi_{-}^{}(0)}_{}^{}$ and $-{\xi_{+}^{}(0)}_{}^{}$ with $+{\xi_{+}^{}(0)}_{}^{}$ (through $\infty$). 
With this choice, we get a single valued function which is analytic in
the strip $|{\xi_{-}^{}(0)}_{}^{}|<|\xi_{}^{}|<|{\xi_{+}^{}(0)}_{}^{}|$. Hence the above complex
integral just gives the Laurent expansion coefficients of
$\sqrt{\frac{\left(1-\frac{1}{{\xi_{+}^{}(0)}_{}^{2}}\right)\left(1-{\xi_{-}^{}(0)}_{}^{2}\right)}{
\left(1-\frac{\xi_{}^{2}}{{\xi_{+}^{}(0)}_{}^{2}}\right)\left(1-\frac{{\xi_{-}^{}(0)}_{}^{2}}{\xi_{}^{2}}\right)}}$ expanded around $\xi_{}^{}=0$. 
The final expression for the distribution is obtained as,  
\begin{eqnarray}
\label{eq-24}
&&P[2 n ; t]=\nonumber\\
&&\left(\frac{1+f}{f}\right)_{}^{n}
\sqrt{\frac{\left(1-\frac{1}{{\xi_{+}^{}(0)}_{}^{2}}\right)\left(1-{\xi_{-}^{}(0)}_{}^{2}\right)}{\pi}} \left(-\frac{{\xi_{-}^{}(0)}_{}^{}}{{\xi_{+}^{}(0)}_{}^{}}\right)_{}^{|n|}
 \nonumber\\
 &&\frac{\Gamma[\frac{1}{2}+|n|]}{\Gamma[1+|n|]}\mathbb{F}_{1}^{2}\left[\frac{1}{2},\frac{1}{2}+|n|,1+|n|;\left(\frac{{\xi_{-}^{}(0)}_{}^{}}{{\xi_{+}^{}(0)}_{}^{}}\right)_{}^{2}\right],
\end{eqnarray}
and 
\begin{eqnarray}
\label{eq-25}
P[2 n + 1; t]=&&0.
\end{eqnarray}
It is important to observe here that there are no microscopic realizations where classical drive does odd quanta of work on the optical oscillator. 
Further, similar to $z_{2}^{}=0$ case, the probability function for work takes same form for both the forward and backward processes. 
The above explicit expression for $P[n ; t]$ satisfies the detailed work fluctuation theorem and hence the integral fluctuation theorem. 

We find that the work distribution is always maximum at $n=0$. 
\cite{Deffner2008,Ford2012}. 
The probability that no work is performed on the optical oscillator is given by, 
\begin{eqnarray}
\label{eq-26}
  P[0 ; t] &=&
\frac{2}{\pi}\sqrt{\left(1-\frac{1}{{\xi_{+}^{}(0)}_{}^{2}}\right)\left(1-{\xi_{-}^{}(0)}_{}^{2}\right)}
K\left[\left(\frac{{\xi_{-}^{}(0)}_{}^{}}{{\xi_{+}^{}(0)}_{}^{}}\right)_{}^{2}\right],\nonumber\\
\end{eqnarray}
where $K[x]$ is the complete elliptic integral of first kind
\cite{Milne1972,Arfken2013}. For the zero temperature case ($\beta\epsilon \to
\infty \Rightarrow f \rightarrow 0$), ${\xi_{-}^{}(0)}_{}^{}=0$ and ${\xi_{+}^{}(0)}_{}^{}=\mbox{coth}^2(|z_{2}^{}|t)$, which makes,
$P[0 ; t]=\mbox{sech}(|z_{2}^{}|t)$ indicating that, for long time, number of
microscopic realizations where 
no work is done on the oscillator are exponentially suppressed. 
Further, in the limit of zero temperature,
\begin{eqnarray}
\label{eq-27}
 P[2 n ; t] &=&
\frac{1}{\sqrt{\pi}}\frac{\Gamma[\frac{1}{2}+n]}{\Gamma[1+n]}\mbox{sech}
(|z_{2}^{}|t)\mbox{tanh}_{}^{2n}(|z_{2}^{}|t) \Theta[n],\nonumber\\
\end{eqnarray}
indicating that there are no microscopic realizations where work is extracted
from the system, this is intuitive, since for zero temperature case, oscillators
initial state is ground state and so it is not possible to extract any work from
it. For large time $|z_{2}^{}|t \rightarrow \infty$, $P[2 n \geq 0; t]$ 
decays as an exponential in time. This behavior is different from $z_{2}^{}=0$ case discussed previously, 
where $P[n \geq 0; t]$ decays as a Gaussian with time.

For large fluctuations (large $n$) probability can be approximated by
\begin{eqnarray}
\label{eq-28}
\bar P[2 n ; t] &=& 
\sqrt{\frac{\left(1-\frac{1}{{\xi_{+}^{}(0)}_{}^{2}}\right)\left(1-{\xi_{-}^{}(0)}_{}^{2}\right)}{
\left(1-\frac{{\xi_{-}^{}(0)}_{}^{2}}{{\xi_{+}^{}(0)}_{}^{2}}\right)}} 
 \frac{e^{- 2 n\mbox{ln}(|{\xi_{\pm}^{}(0)}_{}^{}|)}}{\sqrt{|n|}}.\nonumber\\
\end{eqnarray}
Thus the distribution function falls off exponentially in tails with different
rates determined by 
$|{\xi_{+}^{}(0)}_{}^{}|$ and $|{\xi_{-}^{}(0)}_{}^{}|$. It shows that the large fluctuations in $n$ (or work) are
suppressed exponentially. 

The probability weight for smaller values of $n$ falls quickly as time ($|z_{2}^{}|
t$) increases, however, since ${\xi_{+}^{}(0)}_{}^{}$
approaches to unity, the weight for larger values of $n$ increases.
Further for large times ($|z_{2}^{}| t \to \infty$),
${\xi_{+}^{}(0)}_{}^{} \to 1$ (${\xi_{-}^{}(0)}_{}^{2}{\xi_{+}^{}(0)}_{}^{2} = \left(\frac{f}{1+f}\right)_{}^{2}$
for any time), 
making the distribution function more flatter for positive $n$ with time, but
for negative 'n', tails decay with finite rate even for long time.  

\subsection{$z_{1}^{}\neq 0 \neq z_{2}^{}$ case}

For the general case, the moment generating function of work is given in Eq. (\ref{eq-12}), the first two cumulants of work are given as :
\begin{eqnarray}
\label{eq-29}
\langle n \rangle &=& \left(1+2f\right)\text{sinh}_{}^{2}(|z_{2}^{}|t) + \alpha(\phi,t)
\end{eqnarray}
and 
\begin{eqnarray}
\label{eq-30}
&&\langle \left(n -\langle n \rangle\right)_{}^{2}\rangle = \left(1+\left(1+2f\right)_{}^{2}\text{cosh}(2|z_{2}^{}| t)\right)\text{sinh}^2(|z_{2}^{}| t)\nonumber\\
&+& \left(1+2f\right)\alpha(\phi,t)\frac{\text{cosh}(3|z_{2}^{}|t)+\text{cos}(\phi_{}^{})\text{sinh}(3|z_{2}^{}|t)}
{\text{cosh}(|z_2|t)+\text{cos}(\phi)\text{sinh}(|z_2|t)}.\nonumber\\ 
\end{eqnarray}
It is interesting to note that the $\phi_{}^{}$ independent terms of both the cumulants are the same as that of $z_{1}^{}=0$ case. This is because cumulant generating function is 
sum of the cumulant generating functions for $z_{1}^{}=0$ case and another $\phi_{}^{}$ dependent term. Note that the two contributions to both the cumulants are positive, but for large times $|z_{2}^{}| t \rightarrow \infty$, the second $\phi_{}^{}$ 
dependent terms grow exponentially (the first term always grows exponentially) with time except for the case $\phi=\pi$ where they saturate to a finite value or decay to zero respectively.  

\begin{figure*}[tbh]
\centering
\begin{tabular}{ccc}
\includegraphics[width=6.2cm,height=4.8cm]{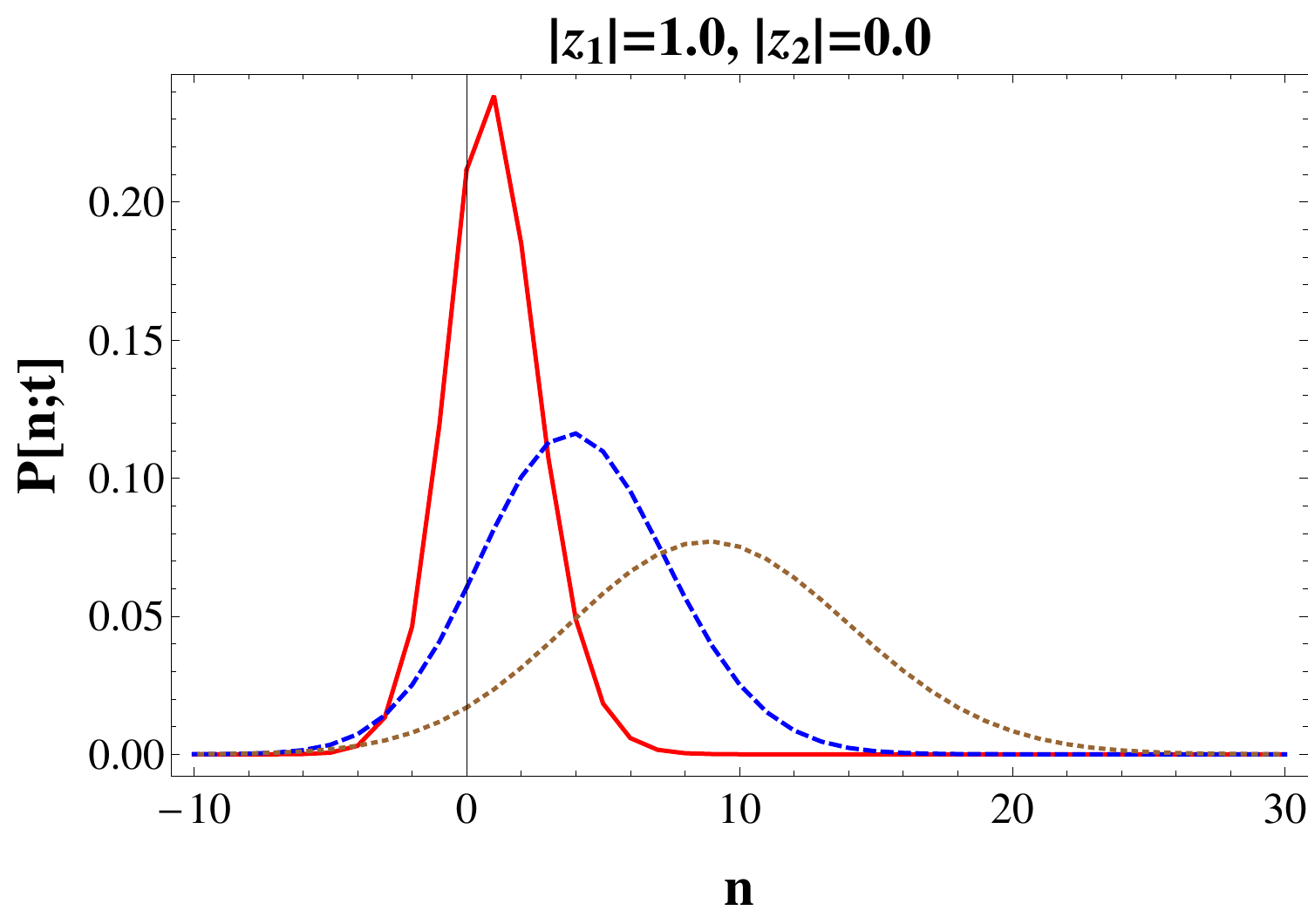}&
&
\includegraphics[width=6.2cm,height=4.8cm]{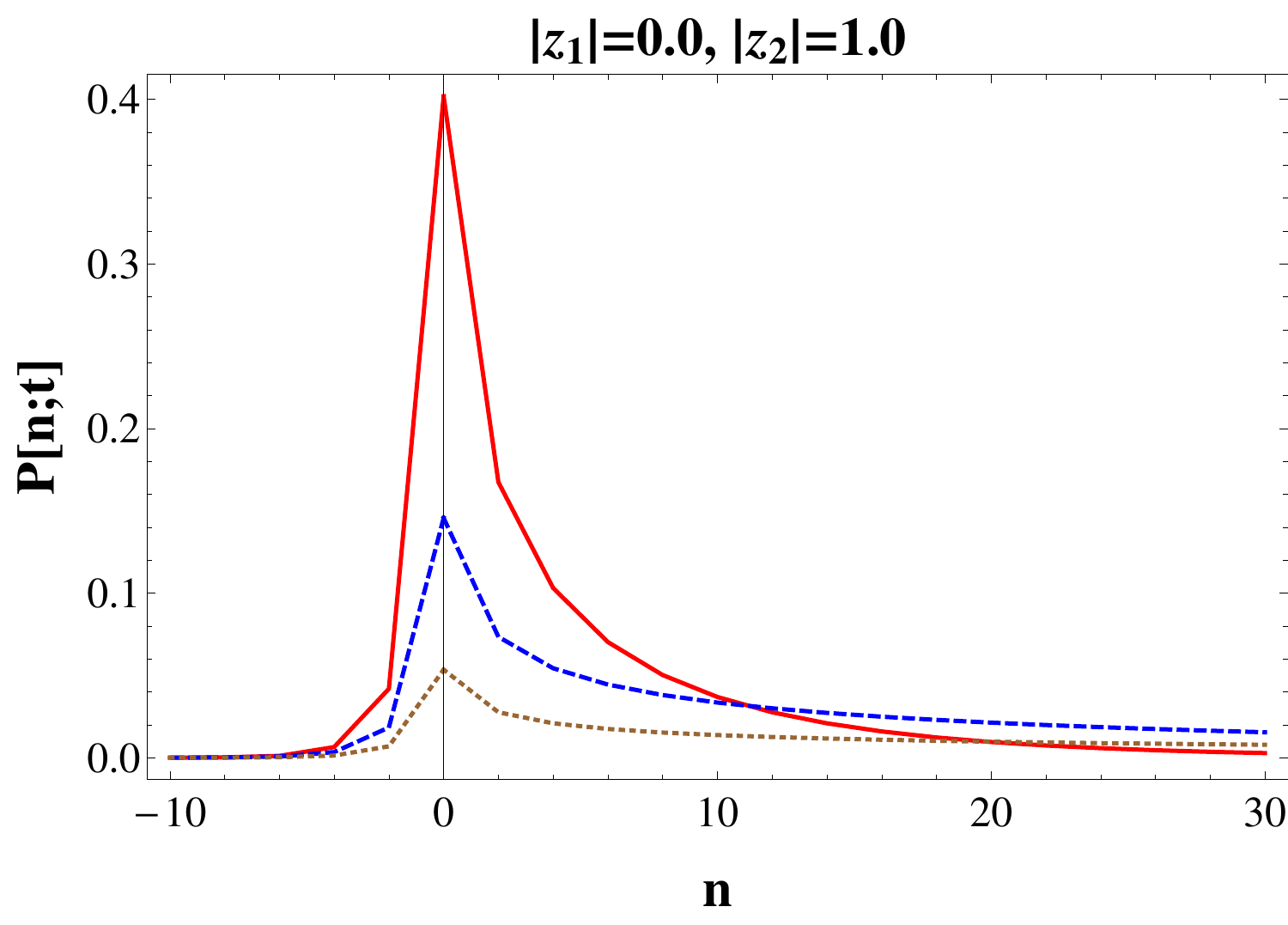}\\ \\
\includegraphics[width=5.5cm,height=4.5cm]{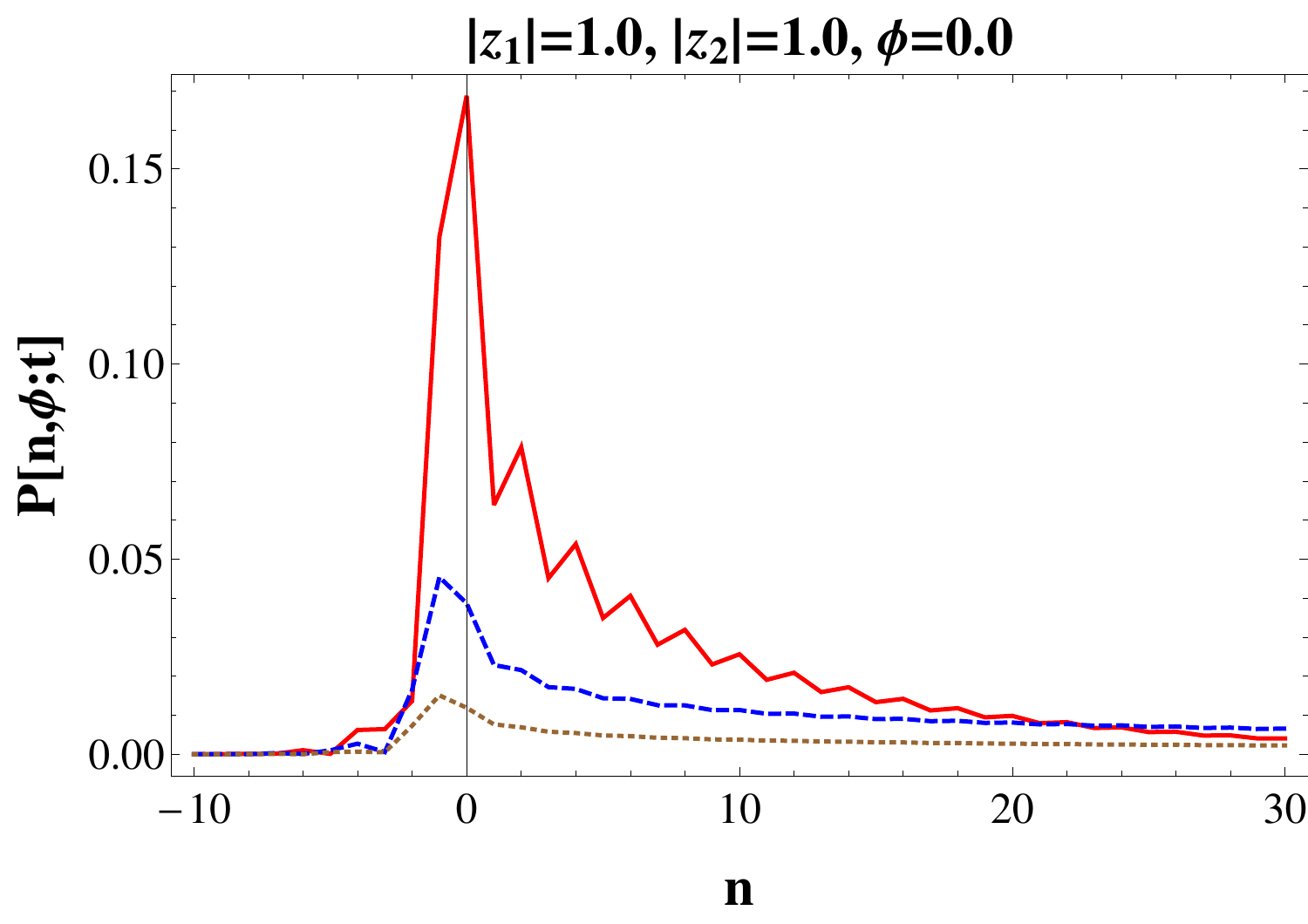}&
\includegraphics[width=5.5cm,height=4.5cm]{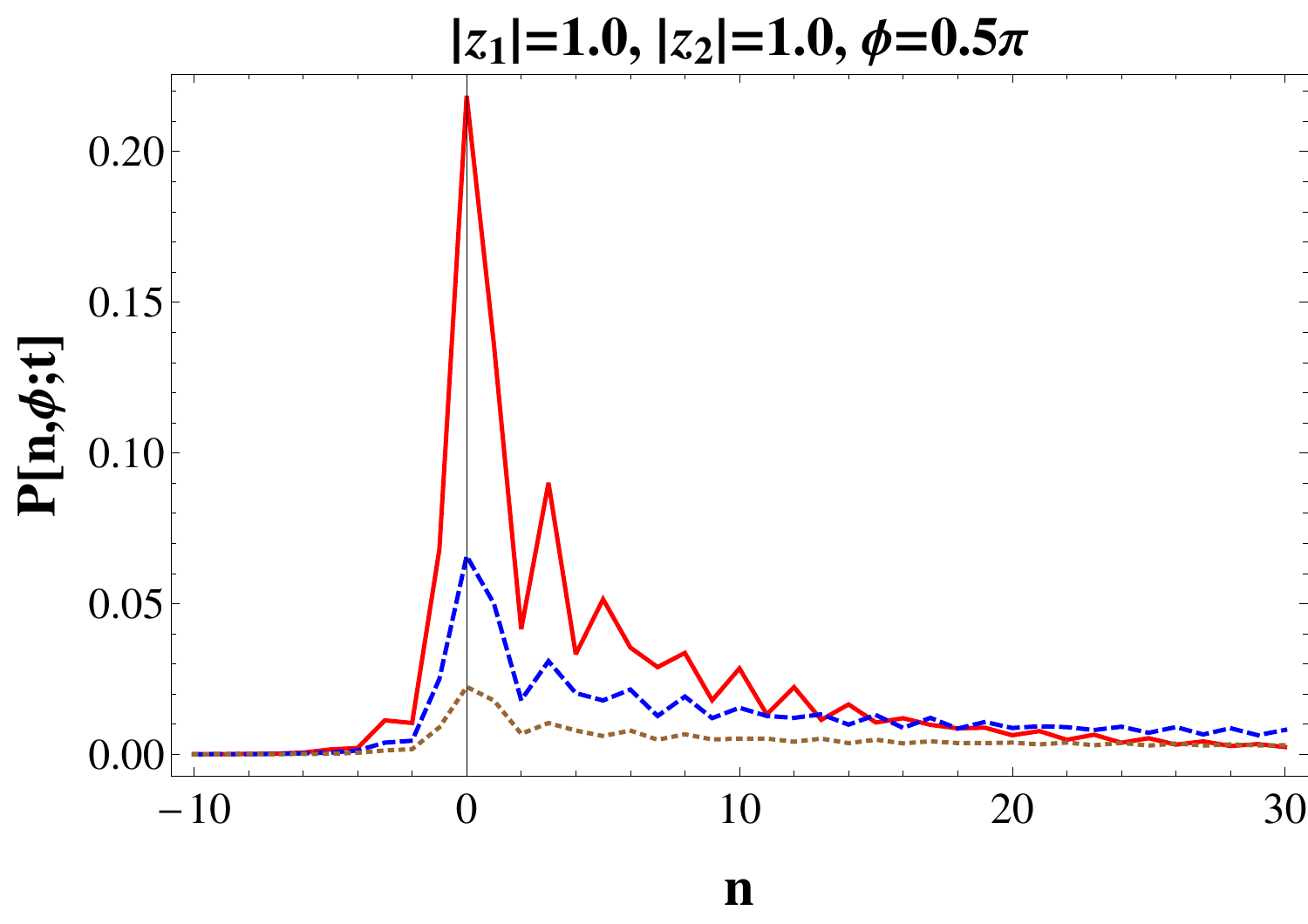}&
\includegraphics[width=5.5cm,height=4.5cm]{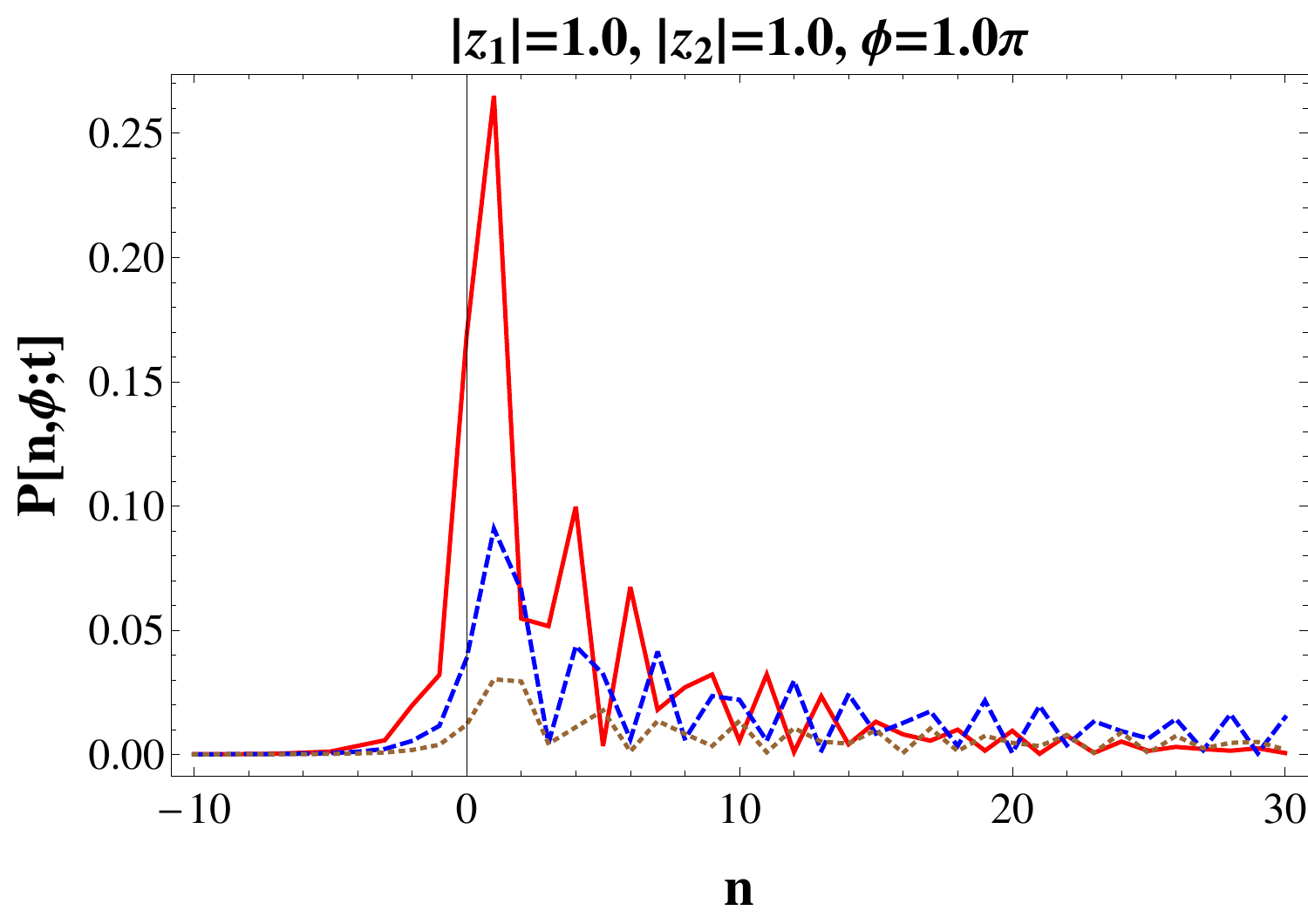}
\end{tabular}
\caption{(Color online) Probability distribution function ($P[n,\phi;t]$) of work done by both the classical drives on the quantum optical oscillator for different measurement times $t=1.0$ (red), $t=2.0$ (dashed-blue), and $t=3.0$ (dotted-brown) with initial (thermal) average photon number $f=1.0$. 
The strength of the classical drives $|z_{1}^{}|$, $|z_{2}^{}|$ and phase difference between them ($\phi_{}^{}$) are indicated in the plots.}
\label{fig-1}
\end{figure*}

We numerically invert the generating function for the general case to obtain probability distribution. 
The probability distribution function for work done $P[n,\phi_{}^{};t]$ obtained analytically for cases $z_{1}^{}=0$ and $z_{2}^{}=0$ and numerically for the general case for different values of $\phi$ at different times 
for fixed values of $|z_{1}^{}|$, $|z_{2}^{}|$ and $f$ is shown in Fig. \ref{fig-1}. 

\begin{figure*}[tbh]
\centering
\begin{tabular}{ccc}
\includegraphics[width=6.0cm,height=5.0cm]{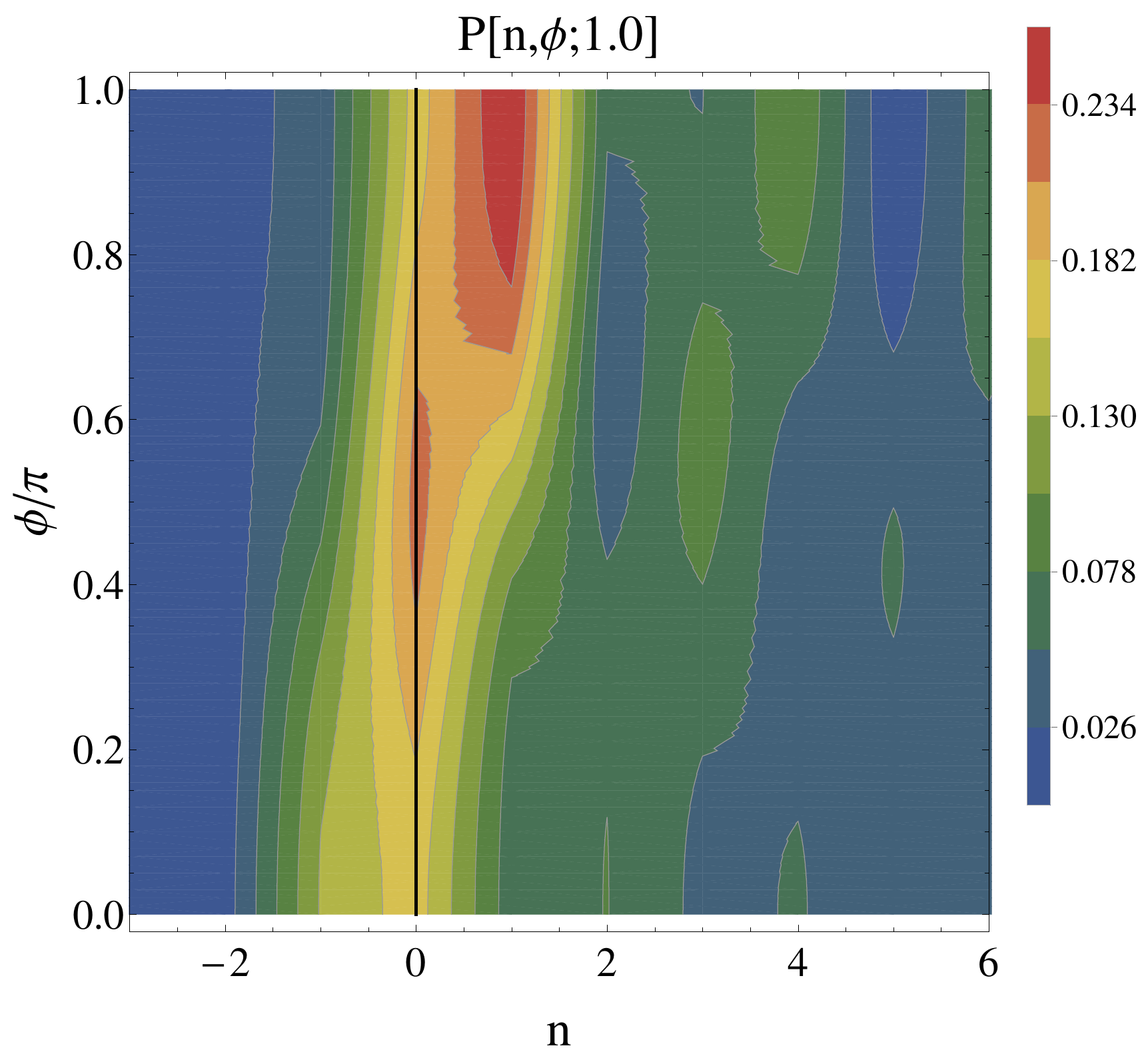}&
\includegraphics[width=6.0cm,height=5.0cm]{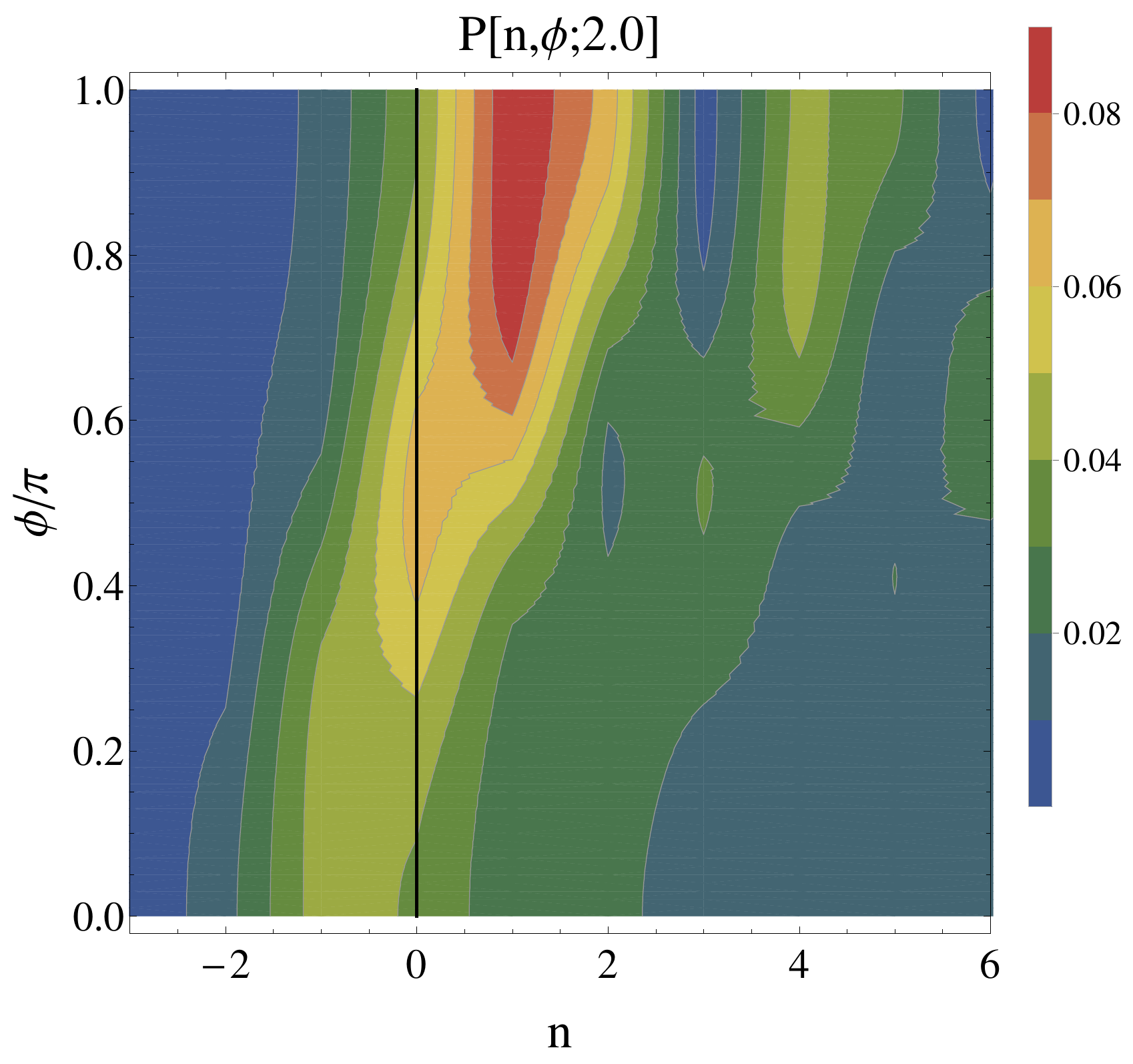}&
\includegraphics[width=6.0cm,height=5.0cm]{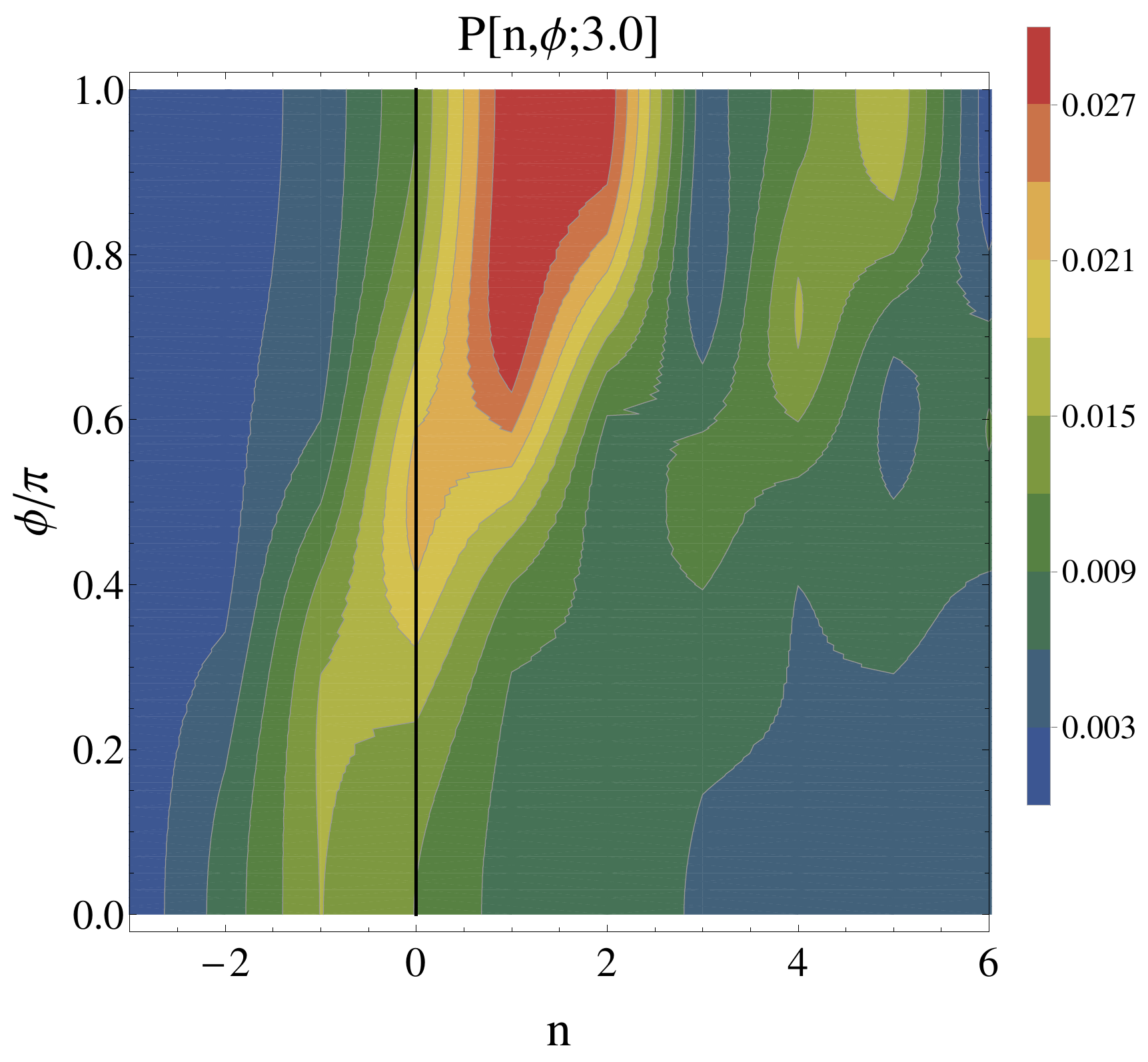}
\end{tabular}
\caption{(Color online) Probability distribution function of work done ($P[n,\phi;t]$) for different measurement times $t=1.0$ (left), $t=2.0$ (center), and $t=3.0$ (right) 
as a function of phase difference between the drives ($\phi_{}^{}$) with $|z_{1}^{}|=|z_{2}^{}|=1.0$ and $f=1.0$.}
\label{fig-2}
\end{figure*}

For $z_{2}^{}=0$ (case A), the distribution is more or less symmetric and the average work roughly corresponds to the peak position which increases quadratically with time. The distribution function behaves 
very differently for $z_{1}^{}=0$ (case B), where the peak of the distribution is always fixed at zero work ($n=0$) while the average work increases exponentially with time 
leading to the asymmetric distribution. For general case (case C), the two drives compete and we find that work distribution becomes more noisy as the value of $\phi$ 
is increased from $0$ to $\pi$. Further for certain values of $\phi$ around 0, the maximum of the distribution function shifts to negative value of $n$ with time as can be seen from Fig. (\ref{fig-2}). 
This is clearly an interference effect between the two drives, for such a behavior is not possible when only one drive is present, as is clear from the analytical results obtained above. 

\section{Conclusion}
\label{conc}

We have computed the statistics of work done by two classical drives, one-photon and two-photon resonant pumps, on the quantum
optical oscillator (a variant of degenerate parametric amplification process). This simple model
allows us to obtain 
exact analytic expression for the moment generating function for work. When only one of the drives is present, the probability function for work is analytically obtained. 
Our results show very different behavior of the work distribution when only individual drivings are present. 
We found that for the case when only one of the pumps are present, work statistics is not influenced by the phase of the drive. When both drives are 
present, the relative phase between the drives influences the work statistics. 
Further for recovering the Jarzynski-Crooks fluctuation theorem, phase has to be reflected around $\pi$ (i.e., $\phi \rightarrow \pi-\phi$) which is 
related to the broken time reversal symmetry of the Hamiltonian. 

\section*{Acknowledgements}
H. Y. and U. H. acknowledge the financial support from the Indian Institute of
Science  (India).

\section*{Appendix}
\subsection{Solution of Eq. (\ref{eq-6})}
Here we present sketch of the method used for solving the parabolic partial differential equation (Eq. (\ref{eq-6})) \cite{Yadalam2019} 
with the initial condition $\mathcal{G}_{\chi}^{}[\zeta_{}^{},\zeta_{}^{*} ; t]|_{t=0}^{}=\mathcal{G}_{}^{}[\zeta_{}^{},\zeta_{}^{*} ;0]$ is given. If $\mathcal{A}_{}^{}=\mathbf{0}$, the above equation is equivalent to the standard Ornstein-Uhlenbeck equation, 
whose solution can be obtained by the method of characteristics \cite{Arfken2013}. For general $\mathcal{A}_{}^{}$, the transformation $\mathcal{G}_{\chi}^{}[\zeta_{}^{},\zeta_{}^{*} ;t]=e_{}^{-\frac{1}{2}\begin{pmatrix} \zeta_{}^{*}  \\ \zeta_{}^{} \end{pmatrix}^{T}_{}\mathcal{U}_{12}^{}(t){\mathcal{U}_{22}(t)}_{}^{-1}\begin{pmatrix} \zeta_{}^{*}  \\ \zeta_{}^{} \end{pmatrix}}\tilde{\mathcal{G}}_{\chi}^{}[\zeta_{}^{},\zeta_{}^{*} ;t]$ 
can be used to eliminate the quadratic term. Here $\mathcal{U}_{xy}^{}(t)$ are square blocks of the following $2 \times 2$ partitioned square matrix,
\begin{equation}
\label{eq-31}
\begin{pmatrix}\mathcal{U}_{11}^{}(t) & \mathcal{U}_{12}^{}(t) \\ \mathcal{U}_{21}^{}(t) & \mathcal{U}_{22}^{}(t)\end{pmatrix}=e_{}^{\begin{pmatrix} \mathcal{B} & -\mathcal{A} \\  \mathcal{C}^{}_{} & -\mathcal{B}_{}^{T} \end{pmatrix}t}.
\end{equation}
With this, $\tilde{\mathcal{G}}_{\chi}^{}[\zeta_{}^{},\zeta_{}^{*} ;t]$ satisfies the following PDE,
\begin{widetext}
\begin{eqnarray}
\label{eq-32}
  &&\frac{\partial}{\partial t} \tilde{\mathcal{G}}_{\chi}^{}[\zeta_{}^{},\zeta_{}^{*} ;
t] =\nonumber\\
&&\left[\frac{1}{2} \begin{pmatrix} \zeta_{}^{*} \\ \zeta_{}^{} \\
\frac{\partial}{\partial \zeta_{}^{*}} \\ \frac{\partial}{\partial \zeta_{}^{}}
\end{pmatrix}_{}^T \begin{pmatrix} \mathbf{0}_{}^{} & \left[\mathcal{B}-\mathcal{U}_{12}^{}(t){\mathcal{U}_{22}(t)}_{}^{-1}\mathcal{C}\right]^{}_{} \\ \left[\mathcal{B}-\mathcal{U}_{12}^{}(t){\mathcal{U}_{22}(t)}_{}^{-1}\mathcal{C}\right]^{T}_{} & \mathcal{C} \end{pmatrix} \begin{pmatrix} \zeta_{}^{*} \\
\zeta_{}^{} \\ \frac{\partial}{\partial \zeta_{}^{*}} \\
\frac{\partial}{\partial \zeta_{}^{}} \end{pmatrix}+\begin{pmatrix} d_{1}^{}-\mathcal{U}_{12}^{}(t){\mathcal{U}_{22}(t)}_{}^{-1} d_{2}^{} \\ d_{2}^{} \end{pmatrix}_{}^{T}\begin{pmatrix} \zeta_{}^{*} \\
\zeta_{}^{} \\ \frac{\partial}{\partial \zeta_{}^{*}} \\
\frac{\partial}{\partial \zeta_{}^{}} \end{pmatrix}\right]\tilde{\mathcal{G}}_{\chi}^{}[\zeta_{}^{},\zeta_{}^{*} ; t],\nonumber\\
 \end{eqnarray}
 \end{widetext}
This PDE can be solved by the method of characteristics to get $\tilde{\mathcal{G}}_{\chi}^{}[\zeta_{}^{},\zeta_{}^{*} ; t]$. Using this, $\mathcal{G}_{\chi}^{}[\zeta_{}^{},\zeta_{}^{*} ; t]$ is given as,
\begin{eqnarray}
\label{eq-33}
 \mathcal{G}_{\chi}^{}[\zeta_{}^{},\zeta_{}^{*} ; t] &=& \int_{\bar\zeta \in 
 \mathbb{C}}^{} d_{}^{2} \bar\zeta \mathbf{G}_{\chi}^{}[\zeta_{}^{} ,
\zeta_{}^{*} 
 ; t | \bar\zeta_{}^{} , \bar\zeta_{}^{*} ; 0 ]
\mathcal{G}_{}^{}[\bar\zeta_{}^{},\bar\zeta_{}^{*} ; 0],\nonumber\\
\end{eqnarray}
with $\mathbf{G}_{\chi}^{}[\zeta_{}^{} , \zeta_{}^{*} ; t | \bar\zeta_{}^{} ,\bar\zeta_{}^{*} ; 0 ]$ given in Eq. (\ref{eq-10}).

\section*{References}
\bibliography{Citation.bib}
\bibliographystyle{unsrt}

\end{document}